\documentstyle[epsfig,12pt]{article}

\newcommand{\ice}[1]{\relax}

\def\bbuildrel#1_#2^#3%
{\mathrel{\mathop{\kern 0pt#1}\limits_{#2}^{#3}}}

\begin{document}
\begin{titlepage}
\noindent
%
%
\hfill TTP94--12
\\
\mbox{}
\hfill  September  1994   \\   %
%
%
\vspace{0.5cm}
\begin{center}
\begin{Large}
\begin{bf}
\renewcommand{\thefootnote}{\fnsymbol{footnote}}
$R_{had}$ at  the $B$-factory\footnote{Work
supported by BMFT contract 056KA93P.
The complete postscript
file of this preprint, including
figures, is available via anonymous ftp at
ttpux2.physik.uni-karlsruhe.de (129.13.102.139) as /ttp94-12/ttp94-12.ps
}
\addtocounter{footnote}{-1}
\renewcommand{\thefootnote}{\arabic{footnote}}
\\
\end{bf}
\end{Large}
%
%
  \vspace{0.8cm}
  \begin{large}
K.G.Chetyrkin$^{a,b}$,
J.H.K\"uhn$^{a}$                      \\[3mm]
   $^a$Institut f\"ur Theoretische Teilchenphysik\\
    Universit\"at Karlsruhe\\
    Kaiserstr. 12,    Postfach 6980\\[1mm]
   D-76128 Karlsruhe, Germany\\[2mm]
$^b$Institute for Nuclear Research \\
Russian Academy of Sciences \\[3mm]
Moscow  117312, Russia
  \end{large}
%
%
  \vspace{1cm}

{\bf Abstract}
\end{center}
\begin{quotation}
\noindent
A systematic theoretical evaluation of the cross section
for hadron production in electron-positron collisions
in the energy range  just below the $B$ meson threshold is
presented which includes charm and bottom mass effects and is
accurate to order $\alpha_s^3$. The corresponding   measurement in the
energy region several $GeV$ above the threshold is also discussed.
\end{quotation}
\end{titlepage}



The hadronic decay rate of the $Z$ boson provides one of the most
accurate values for the strong coupling constant $\alpha_s$.
This measurement  is free from uncertainties and ambiguities which
are inherent  in the precise determination of $\alpha_s$
from  shape variables like thrust distribution, energy-energy
correlation or jet multiplicities and which originate from the
hadronization models, ambiguities in the jet definition etc.
With increasing statistics  and precision   at LEP  the uncertainty
in $\alpha_s$  from this measurement alone  can be reduced  to
$\pm 0.002$. It would be highly desirable  to test the evolution
of the strong coupling as predicted by the beta function
through a determination of $\alpha_s$  from essentially the same
observable, however,  at lower energy. The region from several
GeV above charm threshold (corresponding to  the maximal energy of
BEPC around $5.0$ GeV) to just below the $B$ meson
threshold  at around  $10.5$ GeV corresponding to the
``off  resonance" measurements  of CESR seems particularly suited for
this  purpose.  As a  consequence of  the favorable error propagation,
the accuracy in the  measurement (compared to $91$ GeV)
may decrease by  factor of about 3 or even  4 at 10 and 5.6 GeV
respectively, to achieve comparable precision  in $\Lambda_{QCD}$:
\[
\delta \alpha_s(s) = \frac{\alpha_s^2(s)}{\alpha_s^2(M_Z^2)}
 \delta \alpha_s(M_Z^2)
{}.
\]
Most of the results derived in \cite{Sapirstein,Gorishny1}
for massless quarks are applicable also for the case under
consideration. However, two additional complications arise:
\\
(i) Charm quarks effects cannot be ignored  completely
and  should be taken into consideration  through an expansion in the
ratio $m_c^2/s$, employing the results of \cite{ChetKuhn90,ChetKuhn94}
for terms of order $m_c^2/s$  and $m_c^4/s^2$.
\\
(ii) Contributions involving virtual bottom quarks are presented,
starting from order $\alpha_s^2$. Their contribution depends
in a nontrivial manner on $m_b^2/s$. In order $\alpha_s^2$  these
can be calculated in closed  form and are shown to be small.
Estimates for the corresponding contributions of order
$\alpha_s^3$   indicate that these are under control and can be
safely neglected, provided that one works within the correctly
defined effective four quark theory.

Through most of this paper the results will be formulated in the
$\overline{\mbox{MS}}$ scheme \cite{MSbar}  for a
theory with $n_f =4$ effective flavours and with the
corresponding definitions of the coupling constant and  the
quark mass.  The relation to a formulation with
$f = 5$  appropriate for  the measurements  above the
$b \bar b$  threshold will be given at the end
of the paper.

We shall now list the  independent contributions
and their relative importance.  Neglecting for the moment
the masses of the charmed quark and a forteriori
of the $u$, $d$ and $s$ quarks one predicts in order $\alpha_s^3$
\begin{equation}
R_{NS} = \sum_{f = u,d,s,c}  3 Q_f^2
\left[
1 +  \frac{\alpha_s}{\pi}
+ 1.5245
\left(\frac{\alpha_s}{\pi}\right)^2
-11.52033
\left(\frac{\alpha_s}{\pi}\right)^3
\right]
\label{R3ns}
\end{equation}
for the nonsinglet contribution \cite{Sapirstein,Gorishny1}.
The second and the third order
coefficients are
evaluated with $n_f =4$  which means that the bottom
quark loops are absent. In order $\alpha_s^2$ the bottom quark
\ice{ It is already corrected!!!!!!!!!!!!!
this is for n_f =5
\begin{equation}
R_{NS} = \sum_{f = u,d,s,c}  3 Q_f^2
\left[
1 +  \frac{\alpha_s}{\pi}
+ 1.40923
\left(\frac{\alpha_s}{\pi}\right)^2
-12.7670
\left(\frac{\alpha_s}{\pi}\right)^3
\right].
\label{R3ns}
\end{equation}
}
loops can be taken into consideration with heir full mass dependence.
However, as shown in \cite{Kniehl} the leading term of order
$\alpha_s^2/m_b^2$, which has also been calculated in \cite{me93}
in the framework of an effective $n_f =4$ theory, provides
a fairly accurate description even up to  the very  threshold
$s = 4 m_b^2$.
Hence one has to add a correction term
\begin{equation}
\delta R_{m_b} = \sum_{f = u,d,s,c}  3 Q_f^2
\left(\frac{\alpha_s}{\pi}\right)^2 \frac{s}{\bar m_b^2}
\left[
\frac{44}{675}   +  \frac{2}{135} \log \frac{\bar m_b^2}{s}
\right] .
\label{Rmb}
\end{equation}
For the singlet term one obtains \cite{Gorishny1}.
\begin{equation}
R_S =
-\left(\frac{\alpha_s}{\pi}\right)^3
(\sum_{u,d,s,c} Q_f)^2 \, 1.239 =
-0.55091
\left(\frac{\alpha_s}{\pi}\right)^3
{}.
\label{Rs}
\end{equation}
The bottom quark is absent in this sum.
In view of the smallness of the $\alpha_s^2 s/m^2_b$ correction
(even close to the $\bar b b $ threshold!)
also all other terms of ${\cal O}(\alpha_s^3) $  from virtual
$b$ quarks are neglected. In the same spirit
it is legitimate to use  the scale invariant value of
the $b$  quark mass $\bar m_b = \bar m_b(\bar m^2_b)$ as defined in
the 5 quark theory. Starting from
a pole mass of $4.7$  GeV
and using the results derived in
\cite{polemass1,polemass2}
we get
$\bar{m}_b = 4.10 $ GeV if $\alpha_s^{(5)}(M_Z)= 0.120$.
(Values of $m _b = 4.59 \pm 0.04 $ GeV  \cite{Narison94},
$m _b = 4.72 \pm 0.05 $ GeV \cite{Paver}  and $m_b = 4.7 \pm 0.2$ GeV
\cite{DimaBardin} are currently in use for the pole mass.
)

In contrast to the bottom mass
the effects of the charmed  mass can be incorporated
through an expansion in $m_c^2/s$.
Quadratic mass corrections  have been calculated \cite{ChetKuhn90}
up to order $\alpha_s^3$, quartic mass terms  up to order $\alpha_s^2$
\cite{ChetKuhn94}.
Since $m_c^2/s$ is in itself a small expansion parameter,
the order $\alpha_s^2 m_c^4/s^2$ terms should be sufficient
for the present purpose.

The results for  these corrections read
\begin{equation}
\begin{array}{c}
\delta R_{m_c} =
\\
\displaystyle
3\,Q_c^2 \,12\,\frac{m_c^2}{s} \frac{\alpha_s}{\pi}
\left[
1
+
9.097
\frac{\alpha_s}{\pi}
+
53.453
\left(\frac{\alpha_s}{\pi}\right)^2
\right]
\\
\displaystyle
-
3 \sum_{f=u,d,s,c}Q_f^2\frac{m_c^2}{s}
 \left(\frac{\alpha_s}{\pi}\right)^3
   6.476
\\
\displaystyle
+
3\,Q_c^2 \frac{m_c^4}{s^2}
\left[
-6
-22
\frac{\alpha_s}{\pi}
+
\left(
141.329 - \frac{25}{6}\ln(\frac{m_c^2}{s})
\right)
\left(\frac{\alpha_s}{\pi}\right)^2
\right]
\\
\displaystyle
+3 \sum_{f=u,d,s,c} Q_f^2
\frac{m_c^4}{s^2}
\left(\frac{\alpha_s}{\pi}\right)^2
\left[
-0.4749
- \ln \left(\frac{m_c^2}{s}\right)
\right]
\\
\displaystyle
-3\,Q_c^2 \frac{m_c^6}{s^3}
\left[
8
+\frac{16}{27}
\frac{\alpha_s}{\pi}
\left(
6\ln(\frac{m_c^2}{s}) + 155
\right)
\right]\, ,
\end{array}
\label{Rmc}
\end{equation}
where $n_f=4$ has been adopted everywhere.  Note that
terms of order  $\alpha_s^3 m^2_c/s$ are more important than
those of  order  $\alpha_s^2 m^4_c/s^2$
in the whole energy region under consideration.
For completeness also  $m_c^6/s^3$ and   $\alpha_s m_c^6/s^3$
terms are listed, which, however,  are insignificant amd
will be ignored in the numerical analysis.

The charm quark mass is to be taken as $m_c = \bar m_c^{(4)}(s)$
and is to be evaluated in the 4 flavour theory
via the standard RG equation with the initial
value $\bar m_c(\bar m_c) = 1.11 \ \mbox{GeV}$
corresponding to a pole mass of $1.46$  GeV
in the case of  $\alpha_s^{(5)}(M_Z)= 0.120$
\cite{Narison94}.

A  similar line of reasoning  could be pursued for   bottom
mass terms in the  region several GeV above the $B$ meson
threshold.
It has been argued in \cite{ChetKuhn94,Kuhn94}
(~see also \cite{Soper94})
that mass effects  of order  $\alpha_s$  are very well
parametrized  by taking leading terms of the expansion
in $m_q^2 /s$. This holds  true not only in the
high energy region  but even a few GeV above the threshold.
With this  motivation in mind the terms of order  $\alpha_s^2 m_b^2/s$
and $\alpha_s m_b^6/s^3$  will be   included in the
formula below. The  result is expected to provide a reliable answer
for $\sqrt{s}$ around  15 GeV and perhaps even down to 13 GeV.
The corresponding formula reads as follows
\begin{equation}
\begin{array}{c}
\delta R_{m} =
\\
\displaystyle
3(Q_c^2\frac{m_c^2}{s} + Q_b^2\frac{m_b^2}{s})
12\frac{\alpha_s^{(5)}}{\pi}
\left[
1
+
8.736
\frac{\alpha_s^{(5)}}{\pi}
+
45.657
\left(\frac{\alpha_s^{(5)}}{\pi}\right)^2
\right]
\\
\displaystyle
-3\sum_{f=u,d,s,c,b}Q_f^2
(\frac{m_c^2}{s} +\frac{m_b^2}{s}) \left(\frac{\alpha_s^{(5)}}{\pi}\right)^3
6.126
\\
\displaystyle
+
3\,Q_c^2 \frac{m_c^4}{s^2}
\left[
-6
-22
\frac{\alpha_s^{(5)}}{\pi}
+
\left(
139.488 - \frac{23}{6} \ln(\frac{m_c^2}{s})
+ 12\frac{m_b^2}{m_c^2}
\right)
\left(\frac{\alpha_s^{(5)}}{\pi}\right)^2
\right]
\\
\displaystyle
+
3\,Q_b^2 \frac{m_b^4}{s^2}
\left[
-6
-22
\frac{\alpha_s^{(5)}}{\pi}
+
\left(
139.488 - \frac{23}{6} \ln(\frac{m_b^2}{s})
+ 12\frac{m_c^2}{m_b^2}
\right)
\left(\frac{\alpha_s^{(5)}}{\pi}\right)^2
\right]
\\
\displaystyle
+3\sum_{f=u,d,s,c,b} Q_f^2
\frac{m_c^4}{s^2}
\left(\frac{\alpha_s^{(5)}}{\pi}\right)^2
\left[
-0.4749
- \ln \left(\frac{m_c^2}{s}\right)
\right]
\\
\displaystyle
+3\sum_{f=u,d,s,c,b} Q_f^2
\frac{m_b^4}{s^2}
\left(\frac{\alpha_s^{(5)}}{\pi}\right)^2
\left[
-0.4749
- \ln \left(\frac{m_b^2}{s}\right)
\right]
\\
\displaystyle
\ice{
-3\,Q_c^2 \frac{m_c^6}{s^3}
\left[
8
+\frac{16}{27}
\frac{\alpha_s^{(5)}}{\pi}
\left(
6\ln(\frac{m_c^2}{s}) + 155
\right)
\right]
}
-3\,Q_b^2 \frac{m_b^6}{s^3}
\left[
8
+\frac{16}{27}
\frac{\alpha_s^{(5)}}{\pi}
\left(
6\ln(\frac{m_b^2}{s}) + 155
\right)
\right]\,.
\end{array}
\label{Rm}
\end{equation}
Above the $B$ meson threshold it is more conveniently to express all
quantities for  $n_f = 5 $ theory and thus in (\ref{Rm})
all the coupling constant and quark masses  are evaluated
in the 5 flavour theory at the scale  $\mu = \sqrt{s}$.

The transition from the 4 to
5 flavour theory is performed as follows:
The  charm mass is naturally defined
in the  $n_f =4 $ theory. In order to obtain  the value of
$m_c =\bar m_c^{(5)}(s)$  the initial value
$\bar m_c^{(4)}(1 \mbox{GeV})$ is evolved   via
the $n_f=4$ RG equation to the point $\mu^2 = m_b^2$
and then from there up to $\mu^2 = s$, however,
now with the $n_f=5$ RG equation.
The bottom mass, on the other
hand,  is naturally  defined in the $n_f = 5$
theory irrespective of the characteristic momentum scale of
the problem under consideration.  As  a numerical value we take
the $\bar{m}_b(s)$ obtained from the scale invariant mass
$\bar{m}_b(\bar{m}_b)$ after running the latter
with the help of the $n_f=5$ RG equation.

Finally,
$\alpha_s^{(4)}$ and $\alpha_s^{(5)}$
are related through the following equation \cite{Bernreuther82b}
\begin{equation}
\frac{\alpha_s^{(4)}(\mu)}{\pi} =
\frac{\alpha_s^{(5)}(\mu)}{\pi}
\left[
1 + \frac{1}{6}x\frac{\alpha_s^{(5)}(\mu)}{\pi}
+
\left(
\frac{x^2}{36} + \frac{11 x}{24}
+
\frac{7}{72}
\right)
\left(\frac{\alpha_s^{(5)}(\mu)}{\pi}\right)^2
\right],
\label{WW}
\end{equation}
with $x = \ln(m_b^2(\mu)/\mu^2)$.
Given $\Lambda_{QCD}^{(5)}$, eq. (\ref{WW})
is employed   to
find  $\Lambda_{QCD}^{(4)}$   by setting $\mu= \bar{m_b}(\bar{m_b})$.

In tables  1 -- 6 the predictions for $R$ are listed  for different
values  of $\alpha_s^{(5)}(M_Z)$ together with the values of
$\alpha_s(s)$ and the running masses.
(Note that our predictions are presented without QED corrections
from the running $\alpha$ and initial state radiation.)
Figure 1 shows the behavior of the  ratio $R(s)$
as a function of energy
below and above the bottom threshold, for
$\alpha_s(M_Z) = 0.120, 0.125$ and $0.130$.
The light quark ($u, d, s, c$)  contribution is
displayed separately also above $10.5$ GeV.
It is evident that the predictions from the 4 and 5 flavour
theories join smoothly.  The additional contribution from the
$b \bar{b}$  channel  is presented  down to $11.5$  GeV,
where resonances start to contribute and the perturbative
treatment necessarily ceases to apply. Evidently the $b \bar{b}$
channel is present with full strength  down to the resonance
region -- an important consequence of QCD corrections.
{}From our discussion it should be  evident  that the theoretical
prediction is well under control. Mass effects are small below
the $b \, \bar{b}$ threshold as well as a  few GeV above.
An experimental test is of prime
importance.

\noindent{\large \bf Acknowledgement}
We would like to thank A. Kwiatkowski
for help in checking some results.
K.G.Chetyrkin would like to
thank the Universit\"at Karlsruhe for a guest professorship.


\newpage
\clearpage

\begin{table}
\label{t1}
\centering
\caption{\protect
Values of
$  \protect \Lambda^{(5)}_{\overline{{\scriptstyle MS}}}, \
\Lambda^{(4)}_{\overline{{\scriptstyle MS}}}, \
\alpha_s^{(4)}(s),\     m_c^{(4)}(s)  \
\mbox{and} \  \bar{m}_b(\bar{m}_b)
$
at $\protect\sqrt{s}= 5.0$ GeV
for different values of
$\alpha_s^{(5)}(M_Z^2)\,.$
}
\begin{tabular}{|r|r|r|r|r|r|}
\hline
$\alpha_s^{(5)}(M_Z^2)$              &
$\Lambda^{(5)}_{\overline{{\scriptstyle MS}}}$       &
$\Lambda^{(4)}_{\overline{{\scriptstyle MS}}}$      &
$\alpha_s^{(4)}(s)$                  &
$m_c^{(4)}(s)$                  &
$\bar{m}_b(\bar{m}_b)$          \\
\hline
0.1150   &
175 MeV &
246 MeV  &
0.202   &
0.824 GeV    &
4.16 GeV  \\
0.1175   &
203 MeV &
280 MeV  &
0.210   &
0.777 GeV    &
4.13 GeV  \\
0.1200   &
233 MeV &
317 MeV  &
0.218   &
0.725 GeV    &
4.10 GeV  \\
0.1225   &
266 MeV &
357 MeV  &
0.227   &
0.666 GeV    &
4.07 GeV  \\
0.1250   &
302 MeV &
399 MeV  &
0.236   &
0.599 GeV    &
4.04 GeV  \\
0.1275   &
341 MeV &
444 MeV  &
0.245   &
0.522 GeV    &
4.01 GeV  \\
0.1300   &
383 MeV &
493 MeV  &
0.255   &
0.433 GeV    &
3.98 GeV  \\
\hline
\end{tabular}
\end{table}

\begin{table}
\label{t2}
\centering
\caption{ \protect Predictions for $R(s)$
at $ \protect \sqrt{s}= 5.0$ GeV;
the contributions to $\delta R_{m_c}  $ are shown
separately for every  power of the quark mass.
}
\begin{tabular}{|r|r|r|r|r|r|r|}
\hline
$\alpha_s^{(5)}(M_Z^2)$              &
$ R_{NS}$                       &
$R_S$                           &
$\delta R_{m^2_c}$              &
$\delta R_{m^4_c}$              &
$\delta R_{m_b} $               &
$       R       $               \\
\hline
0.1150   &
3.558   &
-0.00015   &
0.05   &
-0.0066   &
0.0012   &
3.603   \\
0.1175   &
3.567   &
-0.00016   &
0.047   &
-0.0052   &
0.0013   &
3.610   \\
0.1200   &
3.576   &
-0.00018   &
0.044   &
-0.0040   &
0.0014   &
3.618   \\
0.1225   &
3.586   &
-0.00021   &
0.040   &
-0.0028   &
0.0015   &
3.624   \\
0.1250   &
3.596   &
-0.00023   &
0.034   &
-0.0018   &
0.0017   &
3.630   \\
0.1275   &
3.606   &
-0.00026   &
0.028   &
-0.0011   &
0.0019   &
3.634   \\
0.1300   &
3.617   &
-0.00029   &
0.020   &
-0.0005   &
0.0020   &
3.638   \\ 
\hline
\end{tabular}
\end{table}

\begin{table}
\label{t3}
\centering
\caption{\protect
Values of
$  \protect \Lambda^{(5)}_{\overline{{\scriptstyle MS}}}, \
\Lambda^{(4)}_{\overline{{\scriptstyle MS}}}, \
\alpha_s^{(4)}(s),\     m_c^{(4)}(s)  \
\mbox{and} \  \bar{m}_b(\bar{m}_b)
$
at $\protect\sqrt{s}= 10.5$ GeV
for different values of
$\alpha_s^{(5)}(M_Z^2)$\,.
}
\begin{tabular}{|r|r|r|r|r|r|}
\hline
$\alpha_s^{(5)}(M_Z^2)$              &
$\Lambda^{(5)}_{\overline{{\scriptstyle MS}}}$       &
$\Lambda^{(4)}_{\overline{{\scriptstyle MS}}}$      &
$\alpha_s^{(4)}(s)$                  &
$m_c^{(4)}(s)$                  &
$\bar{m}_b(\bar{m}_b)$          \\
\hline
0.1150   &
175 MeV &
246 MeV  &
0.166        &
0.739 GeV    &
4.16 GeV  \\
0.1175   &
203 MeV &
280 MeV  &
0.171        &
0.694 GeV    &
4.13 GeV  \\
0.1200   &
233 MeV &
317 MeV  &
0.176        &
0.644 GeV    &
4.10 GeV  \\
0.1225   &
266 MeV &
357 MeV  &
0.182        &
0.589 GeV    &
4.07 GeV  \\
0.1250   &
302 MeV &
399 MeV  &
0.187        &
0.527 GeV    &
4.04 GeV  \\
0.1275   &
341 MeV &
444 MeV  &
0.193        &
0.457 GeV    &
4.01 GeV  \\
0.1300   &
383 MeV &
493 MeV  &
0.199        &
0.376 GeV    &
3.98 GeV  \\
\hline
\end{tabular}
\end{table}

\begin{table}
\label{t4}
\centering
\caption{ \protect Predictions for $R(s)$
at $ \protect \sqrt{s}= 10.5$ GeV;
the contributions to $\delta R_{m_c}  $ are shown
separately for every  power of the quark mass.
}
\begin{tabular}{|r|r|r|r|r|r|r|}
\hline
$\alpha_s^{(5)}(M_Z^2)$              &
$ R_{NS}$                       &
$R_S$                           &
$\delta R_{m^2_c}$              &
$\delta R_{m^4_c}$              &
$\delta R_{m_b} $               &
$       R       $               \\
\hline
0.1150   &
3.518   &
-0.000081   &
0.0068   &
-0.00022   &
0.0022   &
3.526   \\
0.1175   &
3.523   &
-0.000089   &
0.0063   &
-0.00017   &
0.0024   &
3.532   \\
0.1200   &
3.530   &
-0.000097   &
0.0057   &
-0.00013   &
0.0026   &
3.538   \\
0.1225   &
3.536   &
-0.00011   &
0.0050   &
-0.000088   &
0.0028   &
3.543   \\
0.1250   &
3.542   &
-0.00012   &
0.0042   &
-0.000056   &
0.0030   &
3.549   \\
0.1275   &
3.548   &
-0.00013   &
0.0033   &
-0.000032   &
0.0032   &
3.555   \\
0.1300   &
3.555   &
-0.00014   &
0.0023   &
-0.000015   &
0.0034   &
3.560   \\
\hline
\end{tabular}
\end{table}

\begin{table}
\label{t5}
\centering
\caption{\protect
Values of
$  \protect \Lambda^{(5)}_{\overline{{\scriptstyle MS}}}, \
\Lambda^{(5)}_{\overline{{\scriptstyle MS}}}, \alpha_s^{(5)}(s),
\ m_c^{(5)}(s) \  \mbox{and} \ m_b^{(5)}(s) $
at $\protect\sqrt{s}= 13$ GeV
for different values of
$\alpha_s^{(5)}(M_Z^2)$\,.
}
\begin{tabular}{|r|r|r|r|r|}
\hline
$\alpha_s^{(5)}(M_Z)$                &
$\Lambda^{(5)}_{MS}$            &
$\alpha_s^{(5)}(s)$                  &
$m_c^{(5)}(s)$                  &
$m_b^{(5)}(s)$                  \\
\hline
0.1150   &
175 MeV &
0.162    &
0.720 GeV       &
3.52 GeV  \\
0.1175   &
203 MeV &
0.167    &
0.675 GeV       &
3.47 GeV  \\
0.1200   &
233 MeV &
0.172    &
0.626 GeV       &
3.42 GeV  \\
0.1225   &
266 MeV &
0.177    &
0.572 GeV       &
3.36 GeV  \\
0.1250   &
302 MeV &
0.183    &
0.511 GeV       &
3.30 GeV  \\
0.1275   &
341 MeV &
0.188    &
0.443 GeV       &
3.23 GeV  \\
0.1300   &
383 MeV &
0.194    &
0.364 GeV       &
3.17 GeV  \\
\hline
\end{tabular}
\end{table}

\begin{table}
\label{t6}
\centering
\caption{ \protect Predictions for $R(s)$
at $ \protect \sqrt{s}= 13$ GeV;
the contributions to $\delta R_{m}  $ are shown
separately for every  power of the quark masses.
}
\begin{tabular}{|r|r|r|r|r|r|r|}
\hline
$\alpha_s^{(5)}(M_Z^2)$              &
$ R_{NS}$                       &
$R_S$                           &
$\delta R_{m^2}$                &
$\delta R_{m^4}$                &
$\delta R_{m^6}$                &
$R            $                 \\
\hline
0.1150   &
3.863   &
-0.000019   &
0.027   &
-0.012   &
-0.0016   &
3.876   \\
0.1175   &
3.869   &
-0.000021   &
0.027   &
-0.011   &
-0.0015   &
3.884   \\
0.1200   &
3.875   &
-0.000023   &
0.027   &
-0.011   &
-0.0014   &
3.891   \\
0.1225   &
3.882   &
-0.000025   &
0.027   &
-0.0099   &
-0.0013   &
3.898   \\
0.1250   &
3.888   &
-0.000027   &
0.027   &
-0.0092   &
-0.0011   &
3.905   \\
0.1275   &
3.895   &
-0.000030   &
0.027   &
-0.0086   &
-0.0010   &
3.912   \\
0.1300   &
3.902   &
-0.000032   &
0.026   &
-0.0079   &
-0.00091   &
3.919   \\
\hline
\end{tabular}
\end{table}

\newpage
\protect\clearpage

\setlength{\unitlength}{1cm}

\begin{figure}
\begin{picture}(21.5,11)(0,3)
\put(0,0){\mbox{\epsfig{file=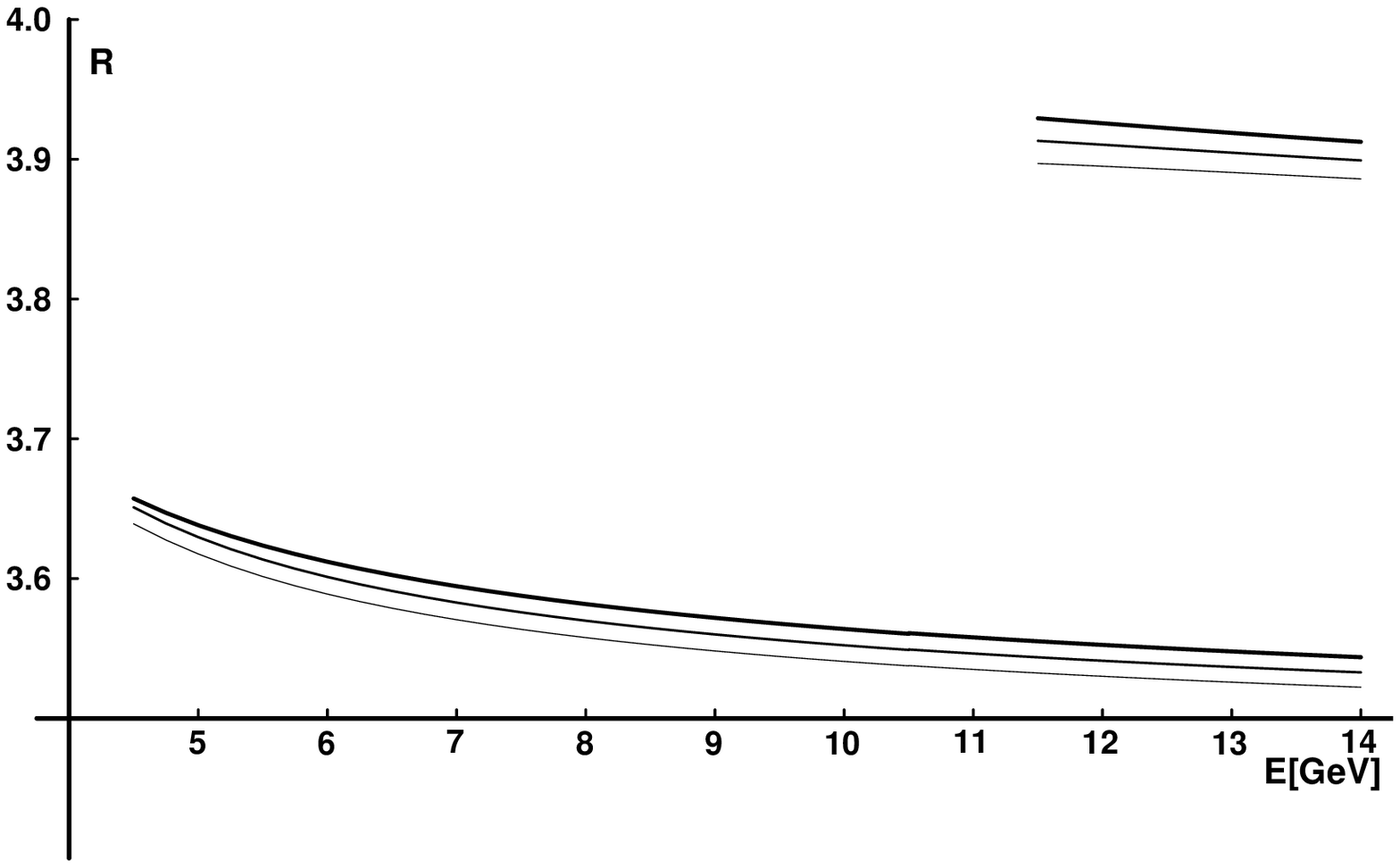,width=14.5cm}}}
\put(0,4){\parbox{14.5cm}{Figure 1: The ratio $R(s)$  below and above
the $b$ quark production threshold at $10.5$ GeV
for  $\alpha_s(M_Z) = 0.120, \, 0.125$ and $0.130$.
The contributions from  light quarks are displayed separately.
}
}
\end{picture}
\end{figure}

\end{document}